\documentclass[epsfig]{emulateapj}

\newcommand{\petroratio}{{{\mathcal{R}}_P}}
\newcommand{\petroradius}{{{\theta}_P}}

\slugcomment{}
\shorttitle{SDSS Pre-Burst Observations of Recent GRB Fields}
\shortauthors{Cool et al. 2005}

\begin{document}
\title{SDSS Pre-Burst Observations of Recent Gamma-Ray
Burst Fields}
\author{Richard J. Cool\altaffilmark{1}, Daniel J.
 Eisenstein\altaffilmark{1}, David W. Hogg\altaffilmark{2},
 Michael R. Blanton\altaffilmark{2}, \\David J. Schlegel\altaffilmark{3},
J. Brinkmann\altaffilmark{4}, Donald P. Schneider\altaffilmark{5},
 Daniel E. Vanden Berk\altaffilmark{5} }

\altaffiltext{1}{Steward Observatory, 933 N Cherry Avenue,
Tucson AZ 85721;
rcool@as.arizona.edu}
\altaffiltext{2}{Center for Cosmology and Particle Physics,
Department of Physics, New York University}
\altaffiltext{3}{Lawrence Berkeley National Laboratory, 
One Cyclotron Road, Mailstop 50R232, Berkeley, CA 94720}
\altaffiltext{4}{Apache Point Observatory, 2001 Apache Point
Road, Sunspot, NM 88349-0059.}
\altaffiltext{5}{Department of Astronomy and Astrophysics,
Pennsylvania State University, 525 Davey Laboratory, University Park,
PA 16802.}

\begin{abstract}

In this paper, we present Sloan Digital Sky Survey (SDSS)
photometry and spectroscopy in the fields of 27 gamma-ray
bursts (GRBs) observed by {\it Swift},
including bursts localized by {\it Swift}, {\it HETE-2},
and {\it INTEGRAL}, after December 2004.  After this
bulk release, we plan to provide individual releases
of similar data shortly after the localization of future
bursts falling in the SDSS survey area.  These data
provide a solid basis for the astrometric and photometric
calibration of follow-up afterglow searches and monitoring.
Furthermore, the images provided with this release
will allow observers to find transient objects up to a
magnitude fainter than possible with Digitized Sky Survey
image comparisons.

\end{abstract}

\keywords{surveys : SDSS, gamma-rays : bursts}

\section{Introduction}

Prompt long wavelength follow-up of gamma-ray bursts
(GRBs) has revolutionized the study of these energetic
and enigmatic objects.  With the successful launch of the
recent high-energy missions such as the {\it High Energy
Transient Explorer 2} \citep[{\it HETE-2}\,;][]{lamb2004},
{\it International Gamma-Ray Astrophysics Laboratory}
\citep[{\it INTEGRAL}\,;][]{winkler2003},and {\it Swift}
\citep{gehrels2004} satellites,  rapid follow-up of
gamma-ray bursts has come to maturity.  {\it Swift} and
{\it HETE-2} not only detect new GRBs but also provide
rapid X-ray localization allowing for very precise 
(uncertainties as small as a few arcseconds) positions, 
far superior than possible in
the past.   The size of  modern GRB samples have become
large enough to study the statistical properties of GRBs
\citep[e.g.][]{berger2005c} and rapid follow-up observations
have allowed for the first optical, radio, and X-ray
localization of afterglows from short-hard gamma-ray bursts
\citep{gehrels2005,hjorth2005,villasenor2005,fox2005,
pedersen2005,prochaska2005,covino2005,berger2005b,bloom2005}.

 There has been large-scale success in identifying
 afterglows and conducting prompt spectroscopic
 follow-up; afterglows extending out to $z\sim6.3$
 \citep[]{price2005,kawai2005}  have been spectroscopically
 confirmed providing luminous probes of the Universe
 even back to the era of reionization.  Recently, high resolution
 spectroscopy of GRBs has allowed for detailed studies
 of absorbing systems arising in the local environment
 of the GRB progenitor \citep{berger2005,hwc2005}.
 Absorption line analyses of GRB afterglow spectra show
 structure similar to that of quasar damped Ly$\alpha$
 Absorbers (DLAs) but extend to higher hydrogen column
 densities and include several metal lines not found in
 quasar DLA systems \citep{watson2005,hwc2005}.

 It has been suggested that GRBs can provide a new class of
 standard candle for future cosmological experiments
 \citep{ghirlanda2004,ghirlanda2004b,
 friedman2005,bloom2003}. As the luminosities
 of these objects should allow detection to very high redshifts
 ($z>10$) \citep{lamb2000},  studies
 of GRBs could extend current work using SN Ia to high
 redshifts, thus providing considerable constraints on our
 understanding of  cosmology.  Before gamma-ray bursts can
 be used as a cosmological tool, however, a large number
 of bursts must be observed spectroscopically in order to
 calibrate the method.

Two keys to studying GRB afterglows are the identification
of a transient afterglow coincident with the burst
detected at high energies and high-quality
information for in-field photometric calibration stars so that
all observations can be placed onto a common photometric system.
Currently, many afterglow searches compare new data to
digitized Palomar Observatory Sky Survey photometric plates
(DSS) and use USNO stars \citep{monnet1998} to calibrate
both astrometry and photometry.  As larger telescopes join
the search for afterglows, the DSS imaging quickly becomes
insufficient, as these images are too shallow for a detailed
comparison to deep optical imaging, so afterglow candidates
can only be identified through their temporal properties.
The USNO catalog is a astrometric catalog, but it was not
intended as a source of photometric calibration across
the sky; USNO-calibrated photometry
 may thus be strongly affected by systematic
photometric residuals in the USNO catalog itself.

The Sloan Digital Sky Survey \citep[SDSS;][]{york2000} provides
accurate
photometry and astrometry for objects over $1/4$ of the
sky, making it a viable alternative to the DSS images
and USNO catalogs. The imaging from SDSS extends over
a magnitude deeper than those from the DSS, making it a
valuable resource in identifying transient objects while
the stable photometric and astrometric calibration of
the survey makes it an ideal source of calibration data
in GRB fields.  In order to aid the community, we are
releasing SDSS imaging and spectroscopy for the fields of
27 {\it Swift}-observed GRBs detected after December 2004,
including bursts originally localized by {\it Swift}, {\it
HETE-2}, and {\it INTEGRAL} observations.  Here, we offer
a bulk release of GRB fields observed in the SDSS, but,
in the future, we will release similar data through the
GRB Coordinate Network (GCN) shortly after the detection
of a new burst within the SDSS survey area.

In this document, we provide some basic documentation
relating to the Sloan Digital Sky Survey and describe the
data products included in each GRB release.  We hope that
the information provided here will be a useful primer
for those unfamiliar with SDSS data, but this document
is by no means intended to be a full description of the
survey or survey data products.  The layout of the paper
is as follows: in \S 2, we describe the Sloan Digital
Sky Survey itself.  A brief description of photometric
quantities measured in SDSS is given in \S 3. We describe
the data products provided with each GRB release in \S4
and offer some concluding remarks in \S5.

\section{The Sloan Digital Sky Survey}
\begin{figure}[!hb]

\centering{\includegraphics[angle=90,
width=3in]{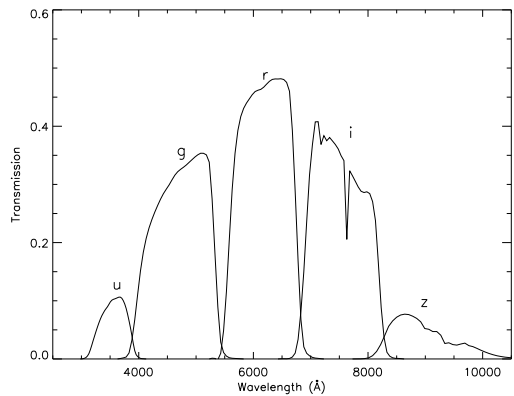}}

\caption{ \scriptsize Transmission curves for the standard
SDSS $ugriz$ photometric system.  Each of the response
curves assumes an air mass of 1.3 and includes the quantum
efficiency of the CCD camera and the reflectivity of the
primary and secondary mirrors on the SDSS telescope.  }
\label{fig:filters}
\end{figure}

The Sloan Digital Sky Survey
\citep{york2000,stoughton2002,a2003,a2004a,a2004b,a2005}
is imaging $\pi$ steradians of the sky through 5
filters, $ugriz$ \citep[see Figure \ref{fig:filters}
and Table 1 for transmission curves and effective
wavelengths]{fukugita1996}.
The imaging is conducted with
a CCD mosaic in drift-scanning mode \citep{gunn1998} on a
dedicated 2.5m telescope \citep{gunn2005} located at Apache
Point Observatory in New Mexico.  Images are processed
\citep{lupton2001,stoughton2002b,pier2003,lupton2005}
and calibrated
\citep{hogg2001,smith2002,ivezic2004,tucker2005}
after which targets are selected for spectroscopy
\citep{eisenstein2001,strauss2002,richards2002} with two
double-spectrographs mounted on the same telescope using a
fiber allocation algorithm which ensures highly complete
spectroscopic samples \citep{blanton2003a}.  The reduced
spectra are classified and redshifts are determined using
the {\tt idlspec2d} automated pipeline (Schlegel et al.,
in preparation).


\begin{deluxetable}{cccccc}
\tablecolumns{6}
\tablenum{1}
\tablewidth{0pt}
\tablecaption{Effective Wavelengths of SDSS Bandpasses}
\tablehead{
\colhead{\bf Filter} & \colhead{$u$} &\colhead{$g$} &
\colhead{$r$} & \colhead{$i$} & \colhead{$z$}}
\startdata
{ $\lambda_{\hbox{eff}} (\mbox{\AA})$} & 3546 & 4670 &
6156 & 7472 & 8917
\enddata
\end{deluxetable}



The photometric calibration of SDSS imaging results in
photometry which is nearly on an AB system \citep{oke1974}.
The details of SDSS photometric calibration are beyond the
scope of this document; the interested reader may find
the calibration procedure in the data release papers
and technical papers cited above; the bottom line is that
the photometric calibration in SDSS is quite robust with
small random errors.    When considering the position of
the bright-star stellar locus in color-color space, the
rms errors in the $g$, $r$,and $i$ bandpasses are found
to be 0.01 mag, 0.02 mag in the $z$-band, and 0.03 in the
$u$-band \citep{ivezic2004}. Similarly small photometric
errors have been demonstrated by measuring the small
scatter in the colors of early-type galaxies on the red
sequence \citep{cool2005}.
It should be noted that the
data provided in these pre-burst GRB releases have been
processed using a slightly different pipeline than that used
for the SDSS public data releases.  We cannot guarantee
that the values presented here will exactly match those
from the SDSS public data releases; in particular, we
expect the photometric calibrations to differ by of order
0.01 mag.

Astrometric calibration within SDSS is also quite
robust.  Relative astrometry is generally better than
50 milli-arcseconds per coordinate while the absolute
astrometry is generally better than 0.1 arcseconds per
coordinate \citep{pier2003}.  It should be noted that the
SDSS astrometric system can have systematic offsets with
respect to other astrometric catalogs. Users requiring high
accuracy astrometry should check for any such offsets when
comparing astrometry between SDSS and data calibrated to
another astrometric system.

Users of $UBVR_CI_C$ imaging may also find the transformations (both
empirically measured and theoretically derived) between the
$UBVR_CI_C$ and
$ugriz$ photometric systems described in \citet{fukugita1996}
and \citet{smith2002}
useful when using the photometry included in our GRB releases to
calibrate new observations.

\section{Measured Quantities}

In this section, we outline the definitions of several of
the photometric measurements used in the SDSS and included
in this data release.  This description is, by no means,
an exhaustive source of information.  Much more detailed
descriptions of the SDSS data products can be found in the
data release papers and the technical papers referenced
in the previous section.


\begin{deluxetable}{cccc}
\tablecolumns{4}
\tablenum{2}
\tablewidth{0pt}
\tablecaption{asinh Magnitude Parameters}
\tablehead{
  \colhead{Filter} &
  \colhead{$b$}&
\colhead{Zero-Flux} &
\colhead{$m_c$} \\
 & & \colhead{Magnitude} & }
\startdata
$u$  & $1.4 \times 10^{-10}$ & 24.63 & 22.12 \\
$g$  & $0.9 \times 10^{-10}$ & 25.11 & 22.60 \\
$r$  & $1.2 \times 10^{-10}$ & 24.80 & 22.29 \\
$i$  & $1.8 \times 10^{-10}$ & 24.36 & 21.85 \\
$z$  & $7.4 \times 10^{-10}$ & 22.83 & 20.32
\enddata
\end{deluxetable}


\subsection{SDSS Magnitudes \& Fluxes}

All SDSS magnitudes, including those presented in this
and future GRB releases, are expressed in terms of asinh
magnitudes \citep{LGS1999}.  At high signal-to-noise,
asinh magnitudes are identical to the standard logarithmic
magnitude \citep{pogson1856}.  Asinh magnitudes, however,
are well behaved even at very low, or even negative,
fluxes, allowing for magnitude calculations even without
a formal object detection.
As described in \citet{stoughton2002}, the asinh magnitude
for a measured flux, $f$, is given by
\begin{equation}
m = - \frac{2.5}{\mbox{ln} 10} \left[\mbox{asinh}\left(
\frac{f/f_0}{2b} \right) + \mbox{ln} b \right] .
\end{equation}
Here $f_0$ defines the zero point of the magnitude
scale and the softening parameter, $b$, denotes the
typical 1 $\sigma$ noise of the sky in a PSF aperture
in 1" seeing.  Table 2, a reproduction of Table 21
in \citet{stoughton2002}, lists the values of $b$ and
the magnitude corresponding to a zero-flux measurement.
The table further lists $m_c$, corresponding to a flux of
$10f_0b$; asinh and logarithmic magnitudes differ by less
than 1\% in flux for objects brighter than this limit.

In this, and future, pre-burst data releases we report
photometry in flux units as well as magnitudes.  All flux
measurements presented in these releases have units
of nanomaggies.  A nanomaggie is a flux-density unit
equal to $10^{-9}$ of a magnitude zero source.  As SDSS is
nearly an AB system, 1 nanomaggie corresponds to $3.631 \,
\mu \mbox{Jy}$ or $3.631 \times 10^{-29}$ erg s$^{-1}$
cm$^{-2}$ Hz$^{-1}$.


\begin{deluxetable}{cl}[b!]
\tablecolumns{2}
\tablenum{3}
\tablewidth{0pt}
\tablecaption{Quantities listed in sdss.calstar files}
\tablehead{
  \colhead{File} &
  \colhead{Description} \\
\colhead{Column} & }
\startdata
1  & Right Ascension (J2000) \\
2  & Declination (J2000)\\
3-7  &  $ugriz$ PSF magnitudes \\
8-12 &  $ugriz$ PSF magnitude errors \\
9-13 &  Quality flags in each filter
\enddata
\end{deluxetable}


\subsection{PSF Magnitudes}
For each object detected in the SDSS imaging, a locally
determined model of the point-spread function (PSF) is
used to measure the flux contained within a PSF centered
on the location of the source. An aperture correction is
applied to each frame based both on the local PSF model
and the seeing in the frame.  Details on these aperture
corrections are given in \citet{stoughton2002}. The
errors reported for the PSF fluxes include contributions
from counting statistics as well as uncertainties in the
aperture corrections and PSF modeling.  PSF magnitudes are
the preferred photometric measurement for point-sources
such as stars and quasars.

\subsection{Model Magnitudes}
Two galaxy models are fit to the two-dimensional image
of each object detected in the SDSS, a pure de Vaucouleurs
\citep{dev1948} profile,
\begin{equation}
I(r) = I_0 \, \mbox{exp} \left(
-7.67[(r/r_{e})^{1/4}]\right)
\end{equation}
and a pure exponential profile
\begin{equation}
I(r) = I_0 \,\mbox{exp}\left( -1.68 r/r_{e} \right) .
\end{equation}
Each of the models are convolved with the local PSF before
fitting to the image.  The best fit model is chosen in the
$r$-band; this is the model used to calculate the model
quantities for the object. In order to provide meaningful
colors, the photometric pipeline fits the full profile
in the $r$-band image and the images from the other bands
are fit allowing only the amplitude of the profile to vary
\citep{stoughton2002}.  In the absence of color-gradients,
model colors provide an unbiased measurement of galaxy
colors.

\subsection{Petrosian Magnitudes}
The Petrosian ratio \citep{petro1976},
$\petroratio$, the ratio of the local surface brightness
at radius $\theta$
to the average surface brightness at that radius, is
given by
\begin{equation}
\label{petroratio}
\petroratio (\theta)\equiv \frac{\left.
\int_{\alpha_{\mathrm{lo}} \theta}^{\alpha_{\mathrm{hi}}
\theta} d\theta' 2\pi \theta'
I(\theta') \right/ \left[\pi(\alpha_{\mathrm{hi}}^2 -
\alpha_{\mathrm{lo}}^2) \theta^2\right]}{\left.
\int_0^\theta dr' 2\pi \theta'
I(\theta') \right/ [\pi \theta^2]},
\end{equation}
where $I(\theta)$ is the azimuthally averaged surface
brightness profile
of a galaxy and $\alpha_{\mathrm{lo}}$ and
$\alpha_{\mathrm{hi}}$ are chosen to
be 0.85 and 1.25 for SDSS. The Petrosian flux is given by
the flux within a circular
aperture of $2 \petroradius$, where $\petroradius$ is the
radius at which
$\petroratio$ falls below 0.2.  In SDSS, $\petroradius$
is determined in
the $r$-band then subsequently used in each of the other
bands.  In the absence of seeing effects, this flux
measurement contains a constant fraction of a galaxy's
light,  independent
of its size or distance, and thus provides a fair flux
measurement when comparing galaxies of different sizes or
at different redshifts. More details of SDSS
Petrosian magnitudes can be found in \citet{blanton2001},
\citet{strauss2002}, and \citet{stoughton2002}.

\subsection{Flags}
A number of data quality flags are maintained in order
to guarantee the accuracy of photometric measurements in
SDSS. A full description of these flags is beyond the scope
of this document but can be found in the Early Data Release
paper \citep{stoughton2002} and at the SDSS DR4 website
\footnote[1]{http://www.sdss.org/dr4/products/catalogs/flags.html}.
For each object included in this release, we provide five
quality flags (one for each band) based on the flags output
by the SDSS pipeline.  In brief, we collapse all of
the SDSS data quality flags
into a single yes or no quality indicator.  Data which
are marked as suspect (flag=1) should be used with caution.

\section{Data Products}
In this section, we describe each of the data products
provided in this data release.  We have removed saturated
stars from all of the data table and we have {\bf not}
corrected the photometry for Galactic extinction.    Again,
magnitudes are asinh magnitudes as is standard in the SDSS.


\begin{deluxetable}{cl}[b]
\tablecolumns{2}
\tablenum{4}
\tablewidth{0pt}
\tablecaption{Quantities listed in sdss.objects files}
\tablehead{
  \colhead{File} &
  \colhead{Description} \\
\colhead{Column} & }
\startdata
1  & Right Ascension (J2000) \\
2  & Declination (J2000)\\
3  & Object type \\
   & star = 6 galaxy = 3 \\
4-8 &  $ugriz$ MODEL magnitudes \\
9-13 &  $ugriz$ MODEL magnitude errors \\
14-18 & $ugriz$ PETROSIAN magnitudes \\
19-23 & $ugriz$ PETROSIAN magnitude errors \\
24-28 & Quality flags in each filter
\enddata
\end{deluxetable}


\subsection{sdss.calstar}

For each burst, we report the photometry and astrometry
of bright stars ($r<20.5$) within 15$'$ of the
burst location in files labeled {\tt sdss.calstar}.
These stars  provide a reliable basis for both the
astrometric and photometric calibration of GRB follow-up
imaging.  The area covered by these calibration stars
is well matched to the {\it Swift}
XRT field-of-view \citep{gehrels2004}.   Table 3 lists
the information provided for each of the calibrations
stars in the {\tt sdss.calstar} file for each GRB dataset.
It is important that the user consult the object flags and
photometric errors when utilizing stars from this file as
some of the stars are poorly detected in the $u$-band.

\subsection{sdss.objects}

In the {\tt sdss.objects\_flux} and {\tt
sdss.objects\_magnitude} files, we provide photometry
and astrometry of all unsaturated objects with $r<23.0$
within 6$'$ of the GRB location.  As suggested
by the filenames, photometric quantities in the {\tt
sdss.objects\_flux} file use flux units (nanomaggies)
while those in the {\tt sdss.objects\_magnitudes} file
use magnitudes (asinh magnitudes).  Table 4 lists the
quantities reported in these files.

We recommend using model fluxes when calculating colors
of galaxies with $r>19.0$. We generally prefer Petrosian
fluxes compared to model fluxes for the overall flux of
a galaxy, but Petrosian fluxes are noisier than model
quantities at faint flux levels.  We warn the user to
exercise caution about using photometry with quoted errors
larger than 20\% in flux.  These objects generally are
simply low signal-to-noise ratio detections, but sometimes
the large errors indicate complications in the reductions.

\begin{figure*}[t]

\centering{\includegraphics[angle=90,
width=6in]{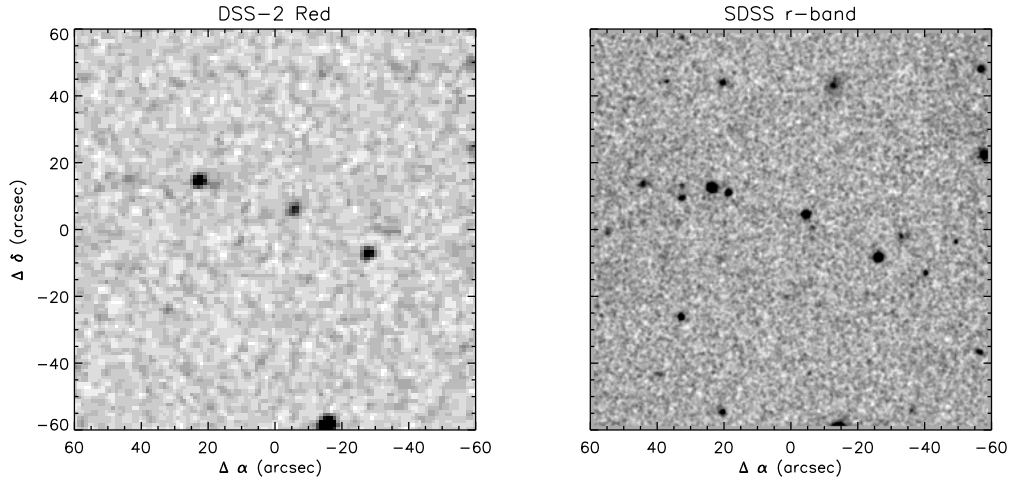}}

\caption{ \scriptsize A comparison of the DSS-2 red
and SDSS imaging for a 1$'$x1$'$ box around the location of
{\it Swift} GRB041224.  The {\it Swift} BAT localization
\citep{barthelmy2004} is at the origin in both images.
The SDSS image clearly extends deeper than the DSS-2 image
making it a more valuable comparison epoch when searching
for GRB afterglows.  }
\label{fig:compare}
\end{figure*}

\subsection{sdss.spectro}

If SDSS spectroscopy has been completed in the field of
the GRB, we include redshifts of spectroscopically observed
objects within 6$'$ of the GRB location in the {\tt
sdss.spectro} file.  Table 5 documents the parameters
reported in the {\tt sdss.spectro} files.  For stars and
galaxies, redshifts are accurate to approximately 30
km s$^{-1}$ while quasar redshifts, which are of less
importance here, are measured to $\Delta z = 0.001$
\citep{stoughton2002}.  All redshifts have been corrected to
a heliocentric frame but are not corrected for Galactic rotation.

Each spectroscopically observed object is classified by
a spectral class (STAR, GALAXY, or QSO) and a spectral
subclass, if appropriate.  Galaxies can be sub-classified
as STARFORMING, AGN, or STARBURST depending on the strength
of the emission lines observed in the spectrum.  Both
galaxies and quasars can be sub-classified as BROADLINE
based on the width of the emission line features; stars
are sub-classified by spectral type.


\begin{deluxetable}{cl}
\tablecolumns{2}
\tablenum{5}
\tablewidth{0pt}
\tablecaption{Quantities listed in sdss.spectro files}
\tablehead{
  \colhead{File} &
  \colhead{Description} \\
\colhead{Column} & }
\startdata
1  & Right Ascension (J2000) \\
2  & Declination (J2000)\\
3  & Spectroscopic classification \\
4  & Spectroscopic subclass  \\
5  & Measured redshift  \\
6  & $1-\sigma$ error on redshift \\
7 & SDSS spectroscopic plate \\
8 & SDSS spectroscopic fiber \\
9 & MJD of SDSS spectroscopic observation
\enddata
\end{deluxetable}


\subsection{Images}

For an 8$'$x8$'$ region around each burst, we include
$g$-zipped FITS images in each of the SDSS passbands
as well as three color-composite ($gri$) images
of the field in JPEG format.   The FITS images are
in units of nanomaggies per pixel where a pixel is
0.396$''$ on a side.  Each of the images is
oriented with N up and E to the left and the FITS headers
include the relevant World Coordinate System information.
The SDSS images provide an important comparison epoch for
follow-up imaging in order to locate possible gamma-ray
burst afterglows.  To illustrate the power of using SDSS
images rather than those from the Digitized Sky Survey,
Figure \ref{fig:compare} shows a 1$'$x1$'$ region around
the field of GRB041224 as imaged in SDSS and the DSS-2
red plates.  Both images are centered on the position
of the GRB as reported by the {\it Swift} BAT detection
\citep{barthelmy2004}.  The imaging from SDSS reaches over
a magnitude deeper than the DSS imaging allowing for the
identification of transient objects to fainter limits than
possible using the DSS.


\begin{deluxetable*}{lrrrrccc}
\tablecolumns{8}
\tablenum{6}
\tablewidth{0pt}
\tablecaption{GRBs covered in this release}
\tablehead{
  \colhead{GRB} &
  \colhead{$\alpha (J2000)$} &
  \colhead{$\delta (J2000)$} &
  \colhead{$N_{\hbox{cal}}$} &
  \colhead{$N_{\hbox{obj}}$} &
  \colhead{$N_{\hbox{spec}}$} &
  \colhead{Position} &
  \colhead{SDSS Status}\\
 & \colhead{(degrees)} & \colhead{(degrees)} &  & & & \colhead{GCN}
 & }
  \startdata
041224     &   56.200 &   $-$6.620 &    280 &    791 &  12 & 2908 & PUBLIC \\ 
050124     &  192.877 &   13.044 &    338 &    688 &   4 & 2974 & PUBLIC \\ 
050215B    &  174.449 &   40.796 &    167 &    463 &   5 & 3027 & PRIVATE \\ 
050319     &  154.200 &   43.548 &    238 &    476 &   2 & 3133 & PRIVATE \\ 
050408     &  180.573 &   10.852 &    231 &    593 &   6 & 3191 & PRIVATE \\ 
050412     &  181.105 &   $-$1.201 &    274 &    450 &   2 & 3241 & PRIVATE \\ 
050416A    &  188.478 &   21.057 &    232 &    596 &   0 & 3268 & PUBLIC \\ 
050504     &  201.005 &   40.703 &    192 &    654 &   1 & 3359 & PRIVATE \\ 
050505     &  141.763 &   30.273 &    286 &    832 &   2 & 3365 & PRIVATE \\ 
050509B    &  189.058 &   28.984 &    118 &    976 &   0 & 3381 & PUBLIC \\ 
050520     &  192.526 &   30.451 &    229 &    858 &   0 & 3434 & PUBLIC \\ 
050522     &  200.081 &   24.770 &    132 &    769 &   0 & 3452 & PUBLIC \\ 
050528     &  353.529 &   45.944 &   2019 &    988 &   3 & 3496 & PUBLIC \\ 
050715     &  155.645 &   $-$0.040 &    392 &    958 &   6 & 3621 & PRIVATE \\ 
050802     &  219.275 &   27.786 &    417 &    622 &   0 & 3734 & PRIVATE \\ 
050813     &  241.988 &   11.248 &    840 &    774 &   0 & 3788 & PUBLIC \\ 
050819     &  358.756 &   24.859 &    598 &   1140 &   0 & 3826 & PUBLIC \\ 
050904     &   13.711 &   14.085 &    412 &    638 &   5 & 3910 & PUBLIC \\ 
050922C    &  317.389 &   $-$8.758 &    663 &    774 &   0 & 4013 & PUBLIC \\ 
051008     &  202.872 &   42.100 &    257 &    408 &   4 & 4071 & PRIVATE \\ 
051114     &  226.267 &   60.156 &    222 &    282 &   1 & 4279 & PRIVATE \\ 
051117A    &  228.391 &   30.870 &    399 &    541 &   4 & 4287 & PRIVATE \\ 
051215     &  163.139 &   38.626 &    162 &   1112 &   8 & 4352 & PRIVATE \\ 
051227     &  125.270 &   31.929 &    552 &   1294 &   4 & 4397 & PUBLIC \\ 
060108     &  147.006 &   31.918 &    187 &    630 &   2 & 4444 & PUBLIC \\ 
060121     &  137.488 &   45.675 &    301 &    507 &   3 & 4550 & PUBLIC \\ 
060123     &  179.700 &   45.513 &    166 &    376 &   1 & 4584 & PRIVATE \\ 
\enddata
\end{deluxetable*}


\section{Data Release and Usage}
In this paper, we present SDSS observations of all {\it
Swift}-observed gamma-ray bursts occurring after
December 2004  that lie in the SDSS
imaging footprint.  For convenience, we include data both
for burst fields with photometry included in previous
SDSS data releases as well as bursts for which the data
has not yet been released by the SDSS.  Table 6 lists the bursts
included in this release as well as the current status
of the data included and the number of calibration stars,
surrounding objects, and spectroscopically observed objects
located in each GRB field.

All of the data included in this release as well as
data we release in the future can be found at  {\bf
http://mizar.as.arizona.edu/grb/public/}.   In addition
to all of the data products described in the previous
section, we also include a short text file for each burst.
This short description provides a brief introduction to
the data products and reports the local mean galactic
extinction for each GRB field. In the future, we will
release SDSS imaging and spectroscopy of individual
GRB fields shortly after future bursts are localized.
These future data releases will be announced via a GCN
circular and, unless otherwise noted, follow the same
data release model described in this document. All data
from this release and future pre-burst data releases may
be used freely, but we request that both the most recent
SDSS data release paper \citep[currently ][]{a2005} and this paper
(for
bursts included in the current release) or GCN circular
(for bursts released in the future) be cited.

\section{Acknowledgments}
RJC is funded through a National Science Foundation
Graduate Student Fellowship.  This research has made use of
NASA's Astrophysics Data System Bibliographic Services. The
Second Palomar Observatory Sky Survey (POSS-II) was made
by the California Institute of Technology with funds from
the National Science Foundation, the National Aeronautics
and Space Administration, the National Geographic Society,
the Sloan Foundation, the Samuel Oschin Foundation, and
the Eastman Kodak Corporation.

Funding for the SDSS and SDSS-II has been provided by the Alfred
P. Sloan Foundation, the Participating Institutions, the National
Science Foundation, the U.S. Department of Energy, the National
Aeronautics and Space Administration, the Japanese Monbukagakusho,
the Max Planck Society, and the Higher Education Funding Council
for England. The SDSS Web Site is http://www.sdss.org/.

The SDSS is managed by the Astrophysical Research Consortium for the
Participating Institutions. The Participating Institutions are the
American Museum of Natural History, Astrophysical Institute Potsdam,
University of Basel, Cambridge University, Case Western Reserve
University, University of Chicago, Drexel University, Fermilab, the
Institute for Advanced Study, the Japan Participation Group, Johns
Hopkins University, the Joint Institute for Nuclear Astrophysics,
the Kavli Institute for Particle Astrophysics and Cosmology, the
Korean Scientist Group, the Chinese Academy of Sciences (LAMOST), Los
Alamos National Laboratory, the Max-Planck-Institute for Astronomy
(MPA), the Max-Planck-Institute for Astrophysics (MPIA), New Mexico
State University, Ohio State University, University of Pittsburgh,
University of Portsmouth, Princeton University, the United States
Naval Observatory, and the University of Washington.

\end{document}